\documentclass{article}
\usepackage[dvips]{graphicx}  

\begin{document}

\begin{center}
{\Large\bf  Black Holes and Quantum Theory: The Fine Structure Constant Connection
\rule{0pt}{13pt}}\par

\bigskip

Reginald T. Cahill \\ 

{\small\it School of Chemistry, Physics and Earth Sciences, Flinders University,

Adelaide 5001, Australia\rule{0pt}{13pt}}\\

\raisebox{-1pt}{\footnotesize E-mail: Reg.Cahill@flinders.edu.au}\par

\vspace{2mm}

{\it Progress in Physics}, {\bf 4}, 44-50, 2006

\vspace{2mm}

\bigskip\smallskip

{\small\parbox{11cm}{%

The new dynamical theory of space is further confirmed by showing that the effective `black hole' masses  $M_{BH}$ in 19 spherical star systems, from globular clusters to galaxies, with masses $M$,  satisfy the prediction that $M_{BH}=\frac{\alpha}{2}M$, where $\alpha$ is the fine structure constant.   As well  the necessary and unique generalisations of the Schr\"{o}dinger and Dirac equations permit the first derivation of gravity from a deeper theory, showing that gravity is a quantum effect of quantum matter interacting with the dynamical space. As well the necessary  generalisation of Maxwell's equations displays the observed light bending effects.  Finally it is shown from the generalised Dirac equation where the spacetime mathematical formalism, and the accompanying geodesic prescription for matter trajectories, comes from.  The new theory of space is non-local and we see many parallels between this and quantum theory, in addition to the fine structure constant manifesting in both, so supporting the argument that space is a quantum foam system, as implied by the deeper information-theoretic theory known as Process Physics. The  spatial dynamics also provides an explanation for the `dark matter' effect and as well the non-locality of the dynamics provides a mechanism for generating the uniformity of the universe, so explaining the cosmological horizon problem.

\rule[0pt]{0pt}{0pt}}}\bigskip

\end{center}

\section{Introduction\label{section:introduction}}
Physics has had two distinct approaches to space. Newton asserted that space existed, but was non-dynamical and unobservable. Einstein, in contrast,  asserted that space was merely an illusion, a perspective effect in that it is four-dimensional spacetime which is real and dynamical, and that the foliation into space and a geometrical model of time was observer dependent; there was no observer independent space.  Hence also according to Einstein  space was necessarily unobservable. However both approaches have been challenged by the recent discovery that space had been detected again and again over more than 100 years \cite{Book,MMCK,AMGE,MMC,MM,Miller,C5,C6,C7,Torr,DeWitte}, and that the dynamics of space is now established\footnote{At least in the limit of zero vorticity.}.  The key discovery \cite{MMCK}  in 2002 was that the speed of light is anisotropic - that it is $c$ only with respect to space itself, and that the solar system has a large speed of some 400km/s relative to that space, which causes the observed anisotropy\footnote{As recently confirmed in a new experiment \cite{anisotropy}.}  This discovery changes all of physics\footnote{Special Relativity does not require that the speed of light be isotropic, as is usually incorrectly assumed.}.  The problem had been that from the very beginning  the various gas-mode Michelson interferometer experiments to detect this anisotropy had been incorrectly calibrated\footnote{Special relativity effects and the presence of  gas in the light paths both play  critical roles  in determining the calibration. In vacuum mode the interferometer is completely insensitive to absolute motion effects, i.e. to the anisotropy of light.}, and that the small fringe shifts actually seen corresponded to this high speed. As well it has been incorrectly assumed that the success of the Special Relativity formalism requires that the speed of light be isotropic, that an actual 3-space be unobservable. Now because  space is known to exist it must presumably also have a dynamics, and this dynamics has been discovered and  tested by explaining various phenomena such as (i) gravity, (ii) the `dark matter' effect, (iii) the bore hole $g$ anomalies, (iv) novel black holes, (v) light bending and gravitational lensing in general, and so on.  Because space has been overlooked in physics as a dynamical aspect of reality all of the fundamental equations of physics, such as Maxwell's equations, the Schr\"{o}dinger equation, the Dirac equation and so on, all lacked the notion that the phenomena described by these equations were  excitations, of various kinds, of the dynamical space itself.
The generalisation of the  Schr\"{o}dinger equation \cite{Schrod} then gave the first derivation and explanation for gravity: it is a quantum effect in which the wave functions are refracted by the inhomogeneities and time variations of the structured space.    However the most striking discovery is that the internal dynamics of space is determined by the fine structure constant \cite{alpha, DM, galaxies, boreholes}.  In this paper we report further observational evidence for this discovery by using a more extensive collection of `black hole' masses in spherical galaxies and globular clusters\footnote{The generic term `black hole'  is used here to refer to  the presence of a compact  closed event horizon enclosing a spatial in-flow singularity.}.  As well we give a more insightful explanation for the dynamics of space.  We also show how  this quantum-theoretic explanation for gravity  leads to a derivation of the spacetime construct where, we emphasise, this is purely a mathematical construct and not an aspect of reality.  This is important as it explains why the spacetime dynamics appeared to be successful, at least in those cases where the `dark matter' effect was not apparent.  However in general the metric tensor of this induced spacetime does not satisfy the General Relativity (GR) equations.  

\section{Dynamics of Space\label{section:dynamics}}

At a deeper level an information-theoretic approach to modelling reality (Process Physics \cite{Book})  leads to an emergent structured `space'  which is 3-dimensional and dynamic, but where the 3-dimensionality is only approximate, in that if we ignore non-trivial topological aspects of space, then it may be embedded in a 3-dimensional  geometrical manifold.  Here the space is a real existent discrete but fractal network of relationships or connectivities,  but the embedding space is purely a mathematical way of characterising the 3-dimensionality of the network.  This is illustrated in Fig.1. This is not an ether model; that notion involved a duality in that both the ether and the space in which it was embedded were both real.  Now the key point is that how we embed the network in the embedding space is very arbitrary: we could equally well rotate the embedding or use an embedding that has the network translating.  These general requirements  then dictate the minimal dynamics for the actual network, at a phenomenological level.  To see this we assume  at a coarse grained level that the dynamical patterns within the network may be described by a velocity field ${\bf v}({\bf r},t)$, where ${\bf r}$ is the location of a small region in the network according to some arbitrary embedding.  For simplicity we assume here that the global topology of the network   is not significant for the local dynamics, and so we embed in an $E^3$, although a generalisation to an embedding in $S^3$ is straightforward.  The minimal dynamics then follows from the above by writing down the lowest order zero-rank tensors, of dimension $1/t^2 $, that are invariant under translation and rotation, giving\footnote{Note that then, on dimensional grounds,  the  spatial dynamics cannot involve the speed of light $c$, except on the RHS where relativistic effects come into play if the speed of matter relative to the local space becomes large, see \cite{Book}. This has significant implications for the nature and   speed of  so-called `gravitational' waves.}

\begin{figure}[t]
\hspace{45mm}\includegraphics[scale=1.4]{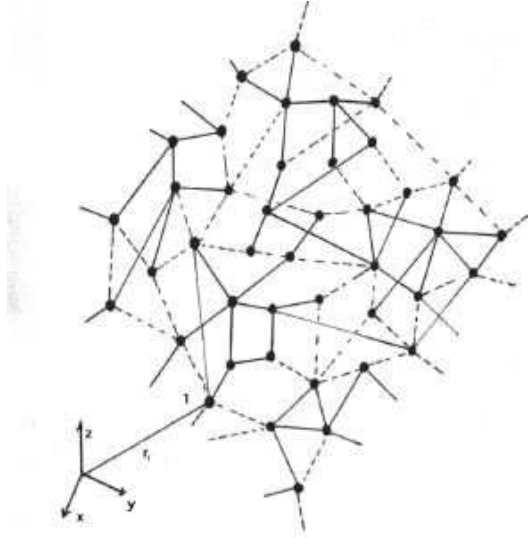}
\caption{\small{  This is an iconic graphical representation of how a dynamical  network has its inherent approximate 3-dimensionality displayed by an embedding in a mathematical   space such as an $E^3$ or an $S^3$.  This space is not real; it is purely a mathematical artifact. Nevertheless this embeddability helps determine the minimal dynamics for the network, as in (\ref{eqn:E1}).  At a deeper level the network is a quantum foam system \cite{Book}.  The dynamical space is not an ether model, as the embedding space does not exist. }
\label{fig:Embedd}}\end{figure}

\begin{equation}
\nabla.\left(\frac{\partial {\bf v} }{\partial t}+({\bf v}.{\bf \nabla}){\bf v}\right)
+\frac{\alpha}{8}(tr D)^2 +\frac{\beta}{8}tr(D^2)=
-4\pi G\rho,
\label{eqn:E1}\end{equation}
where $\rho$ is the effective matter density, and where 
\begin{equation} D_{ij}=\frac{1}{2}\left(\frac{\partial v_i}{\partial x_j}+
\frac{\partial v_j}{\partial x_i}\right).
\label{eqn:E2}\end{equation}

In Process Physics  quantum matter  are topological defects in the network, but here it is sufficient to give a simple description in terms of an  effective density, but which can also model the `dark energy' effect and electromagnetic energy effects, which will be discussed elsewhere. We see that there are only four possible terms, and so we need at most three possible constants to describe the dynamics of space: $G, \alpha$ and $\beta$. $G$ will turn out  to be Newton's gravitational constant, and describes the rate of non-conservative flow of space into matter.  To determine the values of $\alpha$ and $\beta$ we must, at this stage, turn to experimental data.  

However most experimental data involving the dynamics of space is observed by detecting the so-called gravitational  acceleration of matter, although increasingly light bending is giving new information.  Now the acceleration ${\bf a}$ of the dynamical patterns in space is given by the Euler or convective expression
\begin{eqnarray}
{\bf a}({\bf r},t)&\equiv&\lim_{\Delta t \rightarrow 0}\frac{{\bf v}({\bf r}+{\bf v}({\bf r},t)\Delta t,t+\Delta
t)-{\bf v}({\bf r},t)}{\Delta t} \nonumber \\
&=&\frac{\partial {\bf v}}{\partial t}+({\bf v}.\nabla ){\bf v}
\label{eqn:E3}\end{eqnarray} 
and this appears in one of the terms in (\ref{eqn:E1}). As shown in \cite{Schrod} and discussed later herein the acceleration  ${\bf g}$ of quantum matter is identical to this acceleration, apart from vorticity and relativistic effects, and so the gravitational acceleration of matter is also given by (\ref{eqn:E3}).

Outside of a spherically symmetric distribution of matter,  of total mass $M$, we find that one solution of (\ref{eqn:E1}) is the velocity in-flow field  given by\footnote{To see that the flow is inward requires the modelling of the matter by essentially point-like particles. }
\begin{equation}
{\bf v}({\bf r})=-\hat{{\bf r}}\sqrt{\frac{2GM(1+\frac{\alpha}{2}+..)}{r}}
\label{eqn:E4}\end{equation}
but only when $\beta=-\alpha$,  for only then is the acceleration of matter, from (\ref{eqn:E3}), induced by this in-flow of the form
\begin{equation}
{\bf g}({\bf r})=-\hat{{\bf r}}\frac{GM(1+\frac{\alpha}{2}+..)}{r^2}
\label{eqn:E5}\end{equation}
which is Newton's Inverse Square Law of 1687, but with an effective  mass that is different from the actual mass $M$.  So Newton's law requires $\beta=-\alpha$ in (\ref{eqn:E1}) although at present a deeper explanation has not been found.  But we also see modifications coming from the 
$\alpha$-dependent terms.

A major recent discovery \cite{alpha, DM, galaxies, boreholes} has been that experimental data from the bore hole $g$ anomaly has revealed that $\alpha$ is the fine structure constant, to within experimental errors: $\alpha=e^2/\hbar c \approx 1/137.04$. This anomaly is that $g$ does not decrease as rapidly as predicted by Newtonian gravity or GR as we descend down a bore hole.  The dynamics in (\ref{eqn:E1}) and (\ref{eqn:E3}) gives
 the anomaly to be
 \begin{equation}
 \Delta g=2\pi\alpha G \rho d
 \label{eqn:E6}\end{equation}
where $d$ is the depth and $\rho$ is the density, being that of glacial ice in the case of the Greenland Ice Shelf experiments, or that of rock in the Nevada test site experiment. Clearly (\ref{eqn:E6})
permits the value of $\alpha$ to be determined from the data, giving  $\alpha=1/ (137.9 \pm 5)$ from the Greenland Ice Shelf data and, independently, $\alpha=1/ (136.8\pm 3)$ from the Nevada test site data \cite{boreholes}.

In general because (\ref{eqn:E1}) is a scalar equation it is only applicable for vorticity-free flows $\nabla\times{\bf v}={\bf 0}$, for then we can write ${\bf v}=\nabla u$, and then (\ref{eqn:E1}) can always be solved to determine the time evolution of  $u({\bf r},t)$ given an initial form at some time  $t_0$.\footnote{Eqn.(\ref{eqn:E1}) also has Hubble expanding space solutions.}

The $\alpha$-dependent term in (\ref{eqn:E1})  (with now $\beta=-\alpha$) and the matter acceleration effect, now also given by (\ref{eqn:E3}),   permits   (\ref{eqn:E1})   to be written in the form
\begin{equation}
\nabla.{\bf g}=-4\pi G\rho-4\pi G \rho_{DM},
\label{eqn:E7}\end{equation}
where 
\begin{equation}
\rho_{DM}({\bf r},t)\equiv\frac{\alpha}{32\pi G}( (tr D)^2-tr(D^2)),  
\label{eqn:E7b}\end{equation}
where $\rho_{DM}$ is an effective matter density that would be required to mimic the
 $\alpha$-dependent spatial self-interaction dynamics. Then (\ref{eqn:E7}) is the differential form for Newton's law of gravity but with an additional non-matter effective matter density. It has been shown \cite{alpha, DM, galaxies, boreholes}  that this effect explains the so-called `dark matter' effect in spiral galaxies. As shown elsewhere it also explains, when used with the generalised Maxwell's equations,  the gravitational lensing of light by this `dark matter' effect.
 
 An intriguing aspect to the spatial dynamics is that it is non-local.  Historically this was first noticed by Newton who called it action-at-a-distance. To see this we can write  (\ref{eqn:E1}) as an integro-differential equation
 \begin{equation}
 \frac{\partial {\bf v}}{\partial t}=-\nabla\left(\frac{{\bf v}^2}{2}\right)+G\!\!\int d^3r^\prime
 \frac{\rho_{DM}({\bf r}^\prime, t)+\rho({\bf r}^\prime, t)}{|{\bf r}-{\bf r^\prime}|^3}({\bf r}-{\bf r^\prime})
 \label{eqn:E8}\end{equation}
 This shows a high degree of non-locality and non-linearity, and in particular that the behaviour of both $\rho_{DM}$ and $\rho$ manifest at a distance irrespective of the dynamics of the intervening space. This non-local behaviour is analogous to that in quantum systems. 
The non-local dynamics associated with the $\alpha$ dynamics has been tested in various situations, as discussed herein, and so its validity is well established.   This implies that the minimal spatial dynamics in  (\ref{eqn:E1})  involves non-local connectivities.

We term the dynamics of space in ({\ref{eqn:E1}) as a `flowing space'. This term can cause confusion because in normal language a `flow' implies movement of something relative to a background space; but here there is no existent background space, only the non-existent mathematical embedding space.
So here the `flow' refers to internal relative motion, that one parcel of space has a  motion relative to a nearby parcel of space.  Hence the absolute velocities in  ({\ref{eqn:E1})  have no observable meaning; that despite appearances it is only the relative velocities that have any dynamical significance.  Of course it is this requirement that determined the form of ({\ref{eqn:E1}), and as implemented via the embedding space technique.

However there is an additional role for the embedding space, namely as a coordinate system used by a set of cooperating observers. But again while this is useful for their discourse it is not real; it is not part of reality.  

\begin{figure*} 
\hspace{30mm}{\large \begin{tabular}{|c|c|c|c|c|}
\hline\hline
Galaxy & Type & $M_{BH}(+,-)$& $M$&Ref \\
\hline
  M87&E0 & $3.4(1.0,1.0)\times10^9$&$6.2\pm 1.7\times10^{11} $&1\\ 
    NGC4649&E1 & $2.0(0.4,0.6)\times10^9$&$8.4\pm 2.2\times10^{11} $&2\\ 
      M84&E1& $1.0(2.0,0.6)\times10^9$&$5.0\pm 1.4\times10^{11} $&3\\ 
       M32&E2 & $2.5(0.5,0.5)\times10^6$&$9.6\pm 2.6\times10^{8} $&4\\ 
        NGC4697& E4 & $1.7(0.2,0.1)\times10^8$&$2.0\pm 0.5\times10^{11} $&2\\ 
         IC1459&E3 & $1.5(1.0,1.0)\times10^9$&$6.6\pm 1.8\times10^{11} $&5\\ 
          NGC3608&E2 & $1.9(1.0,0.6)\times10^8$&$9.9\pm 2.7\times10^{10} $&2\\ 
           NGC4291&E2& $3.1(0.8,2.3)\times10^8$&$9.5\pm 2.5\times10^{10} $&2\\ 
            NGC3377&E5 & $1.0(0.9,0.1)\times10^8$&$7.8\pm 2.1\times10^{10} $&2\\ 
             NGC4473&E5 & $1.1(0.4,0.8)\times10^8$&$6.9\pm 1.9\times10^{10} $&2\\ 
              CygnusA&E & $2.9(0.7,0.7)\times10^9$&$1.6\pm 1.1\times10^{12} $&6\\ 
               NGC4261&E2 & $5.2(1.0,1.1)\times10^8$&$4.5\pm 1.2\times10^{11} $&7\\ 
                NGC4564&E3 & $5.6(0.3,0.8)\times10^7$&$5.4\pm 1.5\times10^{10} $&2\\ 
                 NGC4742&E4 & $1.4(0.4,0.5)\times10^7$&$1.1\pm 0.3\times10^{10} $&8\\ 
                  NGC3379&E1 & $1.0(0.6,0.5)\times10^8$&$8.5\pm 2.3\times10^{10} $&9\\ 
                   NGC5845&E3 & $2.4(0.4,1.4)\times10^8$&$1.9\pm 0.5\times10^{10} $&2\\ 
                    NGC6251&E2 & $6.1(2.0,2.1)\times10^8$&$6.7\pm 1.8\times10^{11} $&10\\ 
                    \hline\hline
                    Globular  & Cluster  & $M_{BH}(+,-)$& $M$&Ref \\
                    \hline
                       M15&  & $1.7(2.7,1.7)\times10^3$&$4.9 \times10^{5} $&10\\ 
                        G1& & $1.8(1.4,0.8)\times10^4$&$1.35\pm0.5 \times10^{7} $&11\\ 
\hline
 
\hline
 \end{tabular}}
\vspace{3mm}

\hspace{0mm}{Table 1. Black Hole masses and host masses  for various spherical galaxies and globular clusters.   References: 1) Macchetto {\it et al.} 1997; 2) Gebhardt {\it et al.} 2003; 3) average of Bower {\it et al.} 1998;  Maciejewski \& Binney 2001; 4)  Verolme {\it et al.} 2002; 5) average of Verdoes Klein {\it et al.}  2000 and Cappellari {\it et al.}  2002, 6) Tadhunter {\it et al.}  2003; 7) Ferrarese {\it et al.} 1996;  8) Tremaine {\it et al.}  2002;  9) Gebhardt {\it et al.}  2000; 10) Ferrarese \& Ford 1999; 11)  Gerssen {\it et al.} 2002;  12) Gebhardt {\it et al.}  2002.  Least squares best fit of this data to $\mbox{Log}[M_{BH}]=\mbox{Log}[\frac{\alpha}{2}]+x\mbox{Log}[M]$ gives $\alpha=1/ 137.4$ and $x=0.974$. Data and best  fit are shown in Fig.2. Table adapted from Table 1 of \cite{Marconi}.}
\end{figure*}

\section{Black Holes\label{section:blackholes}}

Eqn.(\ref{eqn:E1}) has `black hole' solutions.  The generic term `black hole' is used because they have a compact closed event horizon where the in-flow speed relative to the horizon equals the speed of light, but in other respects they differ from the putative black holes of General Relativity\footnote{It is probably the case that GR has no such solutions - they do not obey the boundary conditions at the singularity, see Crothers \cite{Crothers}.} - in particular their gravitational acceleration is not inverse square law.  The evidence is that it is these new `black holes' from (\ref{eqn:E1}) that have been detected. There are two categories: (i) an in-flow singularity induced by the flow into a matter system, such as, herein, a spherical galaxy or globular cluster. These black holes are termed minimal black holes, as their effective mass is minimal, (ii) primordial naked black holes which then attract matter. These result in spiral galaxies, and the effective mass of the black hole is larger than required merely by the matter induced in-flow. These are therefore termed non-minimal black holes.   These explain the rapid formation of structure in the early universe, as the gravitational acceleration is approximately   $1/r$ rather than $1/r^2$. This is the feature that also explains the so-called `dark matter' effect in spiral galaxies.  Here we consider only the minimal black holes. 

Consider the case where we have a spherically symmetric matter distribution at rest, on average with respect to distant space, and that the in-flow is time-independent and radially symmetric.  Then (\ref{eqn:E1}) is best analysed via (\ref{eqn:E8}),  which can now be written in the form
\begin{equation}\label{eqn:E9}
|{\bf v}({\bf r})|^2=2G\int d^3
r^\prime\frac{\rho_{DM}({\bf r}^\prime)+\rho({\bf r}^\prime)}{|{\bf r}-{\bf r}^\prime|}
\end{equation}
in which the angle integrations may be done to yield
\begin{eqnarray}
v(r)^2&=&\frac{8\pi G}{r}\int_0^r s^2 \left[\rho_{DM}(s)+\rho(s)\right]ds  \nonumber  \\ 
& & +8\pi G\int_r^\infty s
\left[\rho_{DM}(s)+\rho(s)\right]ds, 
\label{eqn:E9a}\end{eqnarray}
 where \footnote{Previous papers had a typo error in this expression. Thanks to Andree Blotz for noting that.}, with $v^\prime=dv(r)/dr$, 
 \begin{equation}
\rho_{DM}(r)= \frac{\alpha}{8\pi G}\left(\frac{v^2}{2r^2}+ \frac{vv^\prime}{r}\right).
\label{eqn:E10}\end{equation}
To obtain the induced in-flow singularity to $O(\alpha)$ we substitute the non-$\alpha$ term in  (\ref{eqn:E9a}) into (\ref{eqn:E10}) giving the effective matter density that mimics the spatial self-interaction of the in-flow,
\begin{equation}
\rho_{DM}(r)=\frac{\alpha}{2r^2}\int_r^\infty s\rho(s)ds+O(\alpha^2).
\label{eqn:E9bb}\end{equation}
We see that  the effective `dark matter' effect is concentrated near the centre, and we find that the total
effective `dark matter' mass is
\begin{eqnarray}
M_{DM} &\equiv& 4\pi\int_0^\infty r^2\rho_{DM}(r)dr  \nonumber \\ &=&\frac{4\pi\alpha}{2}\int_0^\infty
r^2\rho(r)dr+O(\alpha^2) \nonumber \\ &=&\frac{\alpha}{2}M+O(\alpha^2). 
\label{eqn:E10b}\end{eqnarray}
This result applies to any spherically symmetric matter distribution, and is the origin of the $\alpha$  terms in
(\ref{eqn:E4}) and (\ref{eqn:E5}).  It is thus responsible for the bore hole anomaly expression in (\ref{eqn:E6}).
This means that the bore hole anomaly is indicative of an in-flow singularity at the centre of the earth.  This contributes some  0.4\% of the effective mass of the earth, as defined by Newtonian gravity.  However in star systems this minimal black hole effect is more apparent, and we label $M_{DM}$ as $M_{BH}$.  Table 1 shows the effective `black hole' masses attributed to various spherically symmetric star systems based upon observations and analysis of the motion of gases and stars in these systems.  The prediction of the dynamics of space is that these masses should obey (\ref{eqn:E10b}).  The data from Table 1 is plotted in Fig.2, and we see the high precision to which (\ref{eqn:E10b}) is indeed satisfied, and over some 6 orders of magnitude, giving from this data that $\alpha \approx 1/ 137.4$. 

The application of the spatial dynamics to spiral galaxies is discussed in \cite{alpha, DM, galaxies, boreholes} where it is shown that a complete non-matter explanation of the spiral galaxy rotation speed anomaly is given: there is no such stuff as `dark matter' - it is an $\alpha$ determined spatial self-interaction effect.  Essentially even in the non-relativistic regime the Newtonian theory of gravity, with its `universal' Inverse Square Law, is deeply flawed.

\begin{figure*}[t]
\hspace{0mm}\includegraphics[scale=0.5]{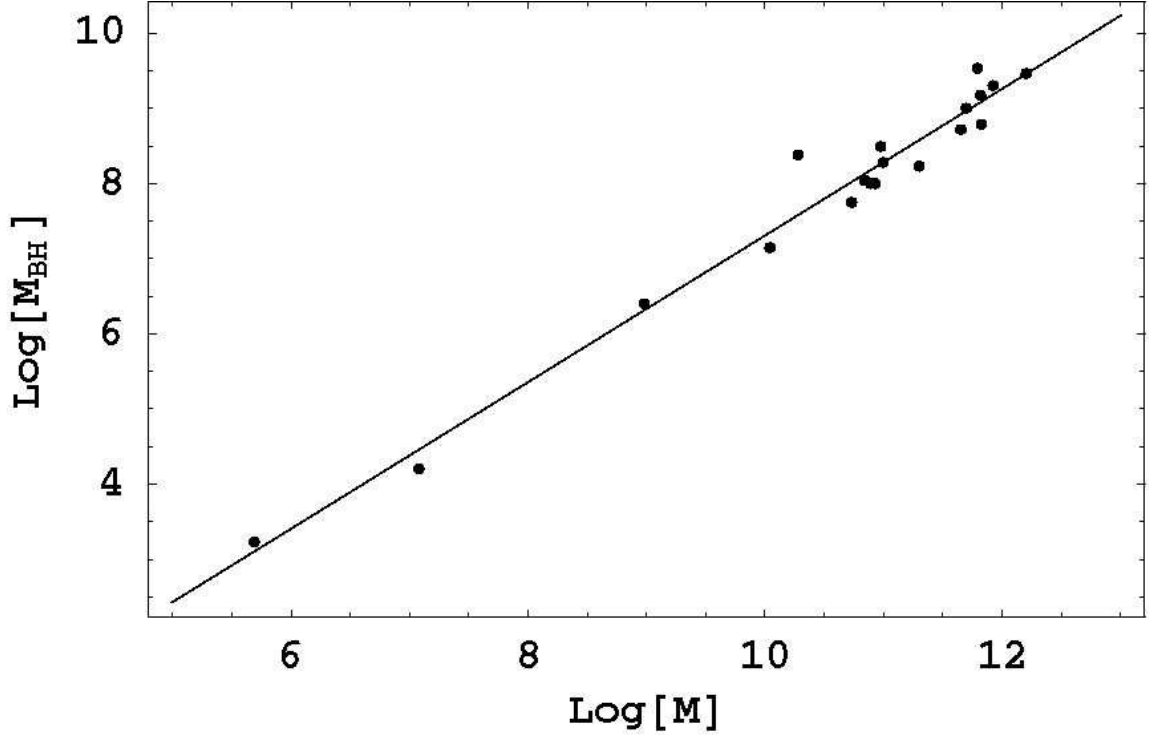}
\caption{ Log-Log  plot of black hole  masses $M_{BH}$ and host galaxy or globular cluster masses $M$ (in solar units) from Table 1.  Straight line is least squares best fit to $\mbox{Log}[M_{BH}]=\mbox{Log}[\frac{\alpha}{2}]+x\mbox{Log}[M]$, giving $\alpha=1/ 137.4$ and $x=0.974$. The borehole $g$-anomaly gives  $\alpha=1/ (137.9 \pm 5)$ from the Greenland Ice Shelf data and $\alpha=1/ (136.8\pm 3)$ from the Nevada test site data \cite{boreholes}. }
\label{fig:BlackHoles}\end{figure*}

\section{Spacetime\label{section:spacetime}}

The curved spacetime explanation for gravity is widely known. Here an explanation for its putative success is given, for there is a natural definition of a spacetime  that arises from (\ref{eqn:E1}), but that it is purely a mathematical construction with no ontological status - it is a mere mathematical artifact.  

First consider the generalised Schr\"{o}dinger \cite{Schrod}
\begin{equation}
i\hbar\frac{\partial  \psi({\bf r},t)}{\partial t}=H(t)\psi({\bf r},t),
\label{eqn:equiv7}\end{equation}
where the free-fall hamiltonian is
\begin{equation}
H(t)=-i\hbar\left({\bf
v}.\nabla+\frac{1}{2}\nabla.{\bf v}\right)-\frac{\hbar^2}{2m}\nabla^2
\label{eqn:equiv8}\end{equation}
As discussed in \cite{Schrod} this is uniquely defined by the requirement that the wave function be attached to the dynamical space, and not to the embedding space, which is  a mere mathematical artifact. We can compute the acceleration of a localised wave packet  according to
\begin{eqnarray}{\bf g}&\equiv&\frac{d^2}{dt^2}\left(\psi(t),{\bf r}\psi(t)\right)   \nonumber \\
&=&\frac{\partial{\bf v}}{\partial t}+({\bf v}.\nabla){\bf v}+
(\nabla\times{\bf v})\times{\bf v}_R
\label{eqn:E11}\end{eqnarray}
where ${\bf v}_R={\bf v}_0-{\bf v}$  is the velocity of the wave packet relative to the local space, as ${\bf v}_0$ is  the velocity relative to the embedding space. Apart from the vorticity term which causes rotation of the wave packet\footnote{This explains the Lense-Thirring effect, and such vorticity  is being detected by the Gravity Probe B satellite gyroscope experiment\cite{GPB}.} we see, as promised, that this matter acceleration is equal to that of the space itself, as in (\ref{eqn:E3}). This is the first derivation of the phenomenon of gravity from a deeper theory: gravity is a quantum effect - namely the refraction of quantum waves by the internal differential motion of the substructure  patterns to space itself. Note that the equivalence principle has now been explained, as this `gravitational' acceleration is independent of the mass $m$ of the quantum system. 

An analogous generalisation of the Dirac equation is also necessary giving the coupling of the spinor to the actual dynamical space, and again not to the embedding space as has been the case up until now, 
\begin{equation}
i\hbar\frac{\partial \psi}{\partial t}=-i\hbar\left(  c{\vec{ \alpha.}}\nabla + {\bf
v}.\nabla+\frac{1}{2}\nabla.{\bf v}  \right)\psi+\beta m c^2\psi
\label{eqn:12}\end{equation}
where $\vec{\alpha}$ and $\beta$ are the usual Dirac matrices. Repeating the analysis in (\ref{eqn:E11}) for the space-induced acceleration we obtain\footnote{Some details are incomplete in this analysis.} 
\begin{equation}\label{eqn:E12}
{\bf g}=\displaystyle{\frac{\partial {\bf v}}{\partial t}}+({\bf v}.{\bf \nabla}){\bf
v}+({\bf \nabla}\times{\bf v})\times{\bf v}_R-\frac{{\bf
v}_R}{1-\displaystyle{\frac{{\bf v}_R^2}{c^2}}}
\frac{1}{2}\frac{d}{dt}\left(\frac{{\bf v}_R^2}{c^2}\right)
\label{eqn:E13a}\end{equation}
which generalises  (\ref{eqn:E11}) by having a term which limits the speed of the wave packet relative to space to be $<\!c$. This equation specifies the trajectory of a spinor wave packet in the dynamical space.  

We shall now show how this leads to both the spacetime mathematical construct and that the geodesic for matter worldlines in that spacetime is equivalent  to trajectories from  (\ref{eqn:E12}).  First we note that (\ref{eqn:E12}) may be obtained by extremising the time-dilated elapsed time 
\begin{equation}
\tau[{\bf r}_0]=\int dt \left(1-\frac{{\bf v}_R^2}{c^2}\right)^{1/2}
\label{eqn:E13}\end{equation}  
with respect to the particle trajectory ${\bf r}_0(t)$ \cite{Book}. This happens because of the Fermat least-time effect for waves: only along the minimal time trajectory do the quantum waves  remain in phase under small variations of the path. This again emphasises  that gravity is a quantum effect.   We now introduce a spacetime mathematical construct according to the metric
\begin{eqnarray}
ds^2&=&dt^2 -(d{\bf r}-{\bf v}({\bf r},t)dt)^2/c^2, \nonumber \\
&=&g_{\mu\nu}dx^{\mu}dx^\nu
\label{eqn:E14}\end{eqnarray}
Then according to this metric the elapsed time in (\ref{eqn:E13}) is
\begin{equation}
\tau=\int dt\sqrt{g_{\mu\nu}\frac{dx^{\mu}}{dt}\frac{dx^{\nu}}{dt}},
\label{eqn:E14b}\end{equation}
and the minimisation of  (\ref{eqn:E14b}) leads to the geodesics of the spacetime, which are thus equivalent to the trajectories from (\ref{eqn:E13}), namely (\ref{eqn:E13a}).
Hence by coupling the Dirac spinor dynamics to the space dynamics we derive the geodesic formalism of General Relativity as a quantum effect, but without reference to the Hilbert-Einstein equations for the induced metric.  Indeed in general the metric of  this induced spacetime will not satisfy  these equations as the dynamical space involves the $\alpha$-dependent  dynamics, and $\alpha$ is missing from GR.   
So why did GR appear to succeed in a number of key tests where the Schwarzschild metric was used?  The answer is provided by identifying the induced spacetime metric corresponding to the in-flow in (\ref{eqn:E4}) outside of a spherical matter system, such as the earth.  Then (\ref{eqn:E14})  becomes
 \begin{eqnarray}
ds^2&=&dt^{ 2}-\frac{1}{c^2}(dr+\sqrt{\frac{2GM(1+\frac{\alpha}{2}+..)}{r}}dt)^2
\nonumber \\ &&-\frac{1}{c^2}r^2(d\theta^{ 2}+\sin^2(\theta)d\phi^2),
\label{eqn:E15}\end{eqnarray}
 Making the change of variables\footnote{No unique choice of variables is required. This choice simply leads to a well-known form for the metric.}
$t\rightarrow t^\prime$ and
$\bf{r}\rightarrow {\bf r}^\prime= {\bf r}$ with

\begin{equation}
t^\prime=t-
\frac{2}{c}\sqrt{\frac{2 GM(1{+}\frac{\alpha}{2}{+}\dots)r}{c^2}}+
+\frac{4\ GM(1{+}\frac{\alpha}{2}{+}\dots)}{c^3}\,\mbox{tanh}^{-1}
\sqrt{\frac{2 GM(1{+}
\frac{\alpha}{2}{+}\dots)}{c^2r}}
\label{eqn:E16}\end{equation}

this becomes (and now dropping the prime notation)
\begin{eqnarray}
ds^2&=&\left(1-\frac{2GM(1+\frac{\alpha}{2}+..)}{c^2r}\right)dt^{ 2} \nonumber \\
&&-\frac{1}{c^2}r^{ 2}(d\theta^2+\sin^2(\theta)d\phi^2) \nonumber \\
&&-\frac{dr^{ 2}}{c^2\left(1-{\displaystyle\frac{
2GM(1+\frac{\alpha}{2}+..)}{ c^2r}}\right)}.
\label{eqn:E17}\end{eqnarray}
which is  one form of the the Schwarzschild metric but with the $\alpha$-dynamics induced effective mass shift. Of course this is only valid outside of the spherical matter distribution, as that is the proviso also on (\ref{eqn:E4}). As well the above particular change of coordinates also introduces spurious singularities at the event horizon\footnote{The event horizon of  (\ref{eqn:E17}) is at a different radius from the actual event horizon of the black hole solutions that arise from   (\ref{eqn:E1})},  but other choices do not do this. 
Hence in the case of the Schwarzschild metric the dynamics missing from both the Newtonian theory of gravity and General Relativity is merely hidden in a mass redefinition, and so didn't affect the various standard tests of GR, or even of Newtonian gravity.  Note that as well we see that the Schwarzschild metric is none other than Newtonian gravity in disguise, except for the mass shift.  While we have now explained why the GR formalism appeared to work, it is also clear that this formalism hides the manifest dynamics of the dynamical space, and which has also been directly detected in gas-mode interferometer and coaxial-cable experiments.

One of the putative key tests of the GR formalism was the gravitational bending of light. This also immediately follows from the new space dynamics once we also generalise the Maxwell equations so that the electric and magnetic  fields are excitations of the dynamical space. The dynamics of the electric and magnetic fields  must then have the form, in empty space,  
\begin{eqnarray}
&&\displaystyle{ \nabla \times {\bf E}=-\mu\left(\frac{\partial {\bf H}}{\partial t}+{\bf v.\nabla H}\right)}\nonumber \\
&&\displaystyle{ \nabla \times {\bf H}=\mbox{\ \ \  }\epsilon\left(\frac{\partial {\bf E}}{\partial t}+{\bf v.\nabla E}\right)} \nonumber \\ 
&&\displaystyle{\nabla.{\bf H}={\bf 0}} , \mbox{\ \ \ \  }
\displaystyle{\nabla.{\bf E}={\bf 0}}
\label{eqn:E18}\end{eqnarray}
which was first suggested by Hertz in 1890  \cite{Hertz}. As discussed elsewhere the speed of EM radiation is now $c=1/\sqrt{\mu\epsilon}$ with respect to the space, and in general not with respect to the observer if the observer is moving through space, as experiment has indicated again and again.
In particular the in-flow in (\ref{eqn:E4}) causes a refraction effect of light passing close to the sun, with the angle of deflection given by
\begin{equation}
\delta=2\frac{v^2}{c^2}=\frac{4GM(1+\frac{\alpha}{2}+..)}{c^2d}
\label{eqn:E19}\end{equation}
where $v$ is the in-flow speed at the surface of the sun, and $d$ is the impact parameter, essentially the radius of the sun. Hence the  observed deflection of $8.4\times10^{-6}$ radians is actually a measure of the in-flow speed at the sun's surface, and that gives $v=615$km/s.  At the earth distance the sun induced spatial  in-flow speed is 42km/s, and this has been extracted from the 1925/26 gas-mode interferometer Miller data \cite{Book,AMGE}. These radial in-flows are to be vectorially summed to the galactic flow of some 400km/s,
but since that flow is much more uniform it does not affect the light bending by the sun in-flow component\footnote{The vector superposition effect for spatial flows is only approximate, and is discussed in more detail in \cite{super}. The solar system has a galactic velocity of some 420$\pm$30km/s in the direction RA=5.2hr, Dec=-67$^0$, as confirmed in a new light-speed anisotropy experiment \cite{anisotropy}.}. Hence the deflection of light by the sun is a way of directly measuring the in-flow speed at the sun's surface, and has nothing to do with `real' curved spacetime. These generalised Maxwell equations also predict gravitational lensing produced by the large in-flows associated with new `black holes' in galaxies.  So again this effect permits the direct observation of the these  black hole effects with their non-inverse square law accelerations.

\section{Conclusions}

We have shown how minimal assumptions about the internal dynamics of space, namely how embeddability in a mathematical space such as an $E^3$ or an $S^3$, expressing its inherent 3-dimensionality,  leads to various predictions ranging from the   anisotropy of the speed of light, as expressed in the required generalisation of Maxwell' s equations, and which has been repeatedly observed since the Michelson-Morley experiment \cite{MM} of 1887, to  the derivation of the phenomenon of gravity that follows after we generalise the Schr\"{o}dinger and Dirac equations.  This shows that the gravitational acceleration of matter is a quantum effect: it follows from the refraction of quantum waves in the inhomogeneities and time-dependencies of the flowing dynamical space. In particular the analysis shows that the acceleration of quantum matter is identical to the convective acceleration of the structured space itself.  This is a non-trivial result.  As well in the case of the Dirac equation we derive the spacetime formalism as well as the geodesic description of matter trajectories, but in doing so reveal that the spacetime is merely a mathematical construct.   We note that the relativistic features of the Dirac equation are consistent with the absolute motion of the wave function in the dynamical 3-space. This emphasis yet again that Special Relativity does not require  isotropy of the speed of light,  as is often incorrectly assumed.

 Here we have further extended the observational evidence that it is the fine structure constant that determines the strength of the spatial self-interaction in this new physics by including data from  black hole masses in 19 spherical star systems. Elsewhere we have already shown that the new space dynamics explains also the spiral galaxy rotation velocity anomaly; that it is not caused by a new form of matter, that the notion of `dark matter' is just a failure of Newtonian gravity and GR.  We have also shown that the space dynamics is non-local, a feature that Newton called action-at-a-distance. This is now extended to include the effects of the spatial self-interaction.  The numerous confirmations of that dynamics, summarised herein, demonstrate the validity of this  non-local physics. Of course since Newton we have become more familiar with non-local effects in the quantum theory.  The new space dynamics shows that non-local effects are more general than just subtle effects in the quantum theory, for in the space dynamics this non-local dynamics is responsible for the supermassive black holes in galaxies. This non-local  dynamics is responsible for two other effects: (i) that the dynamics of space within an event horizon, say enclosing a black hole in-flow singularity  affects the space outside of the horizon, even though EM radiation and matter cannot propagate out through the event horizon, as there the in-flow speed exceeds the speed of light.  So in this new physics we have the escape of information from within the event horizon,  and (ii) that  the universe overall is more highly connected than previously thought. This may explain why the universe is more uniform than expected on the basis of interactions limited by the speed of light, i.e we probably have a solution to the cosmological horizon problem.

Elsewhere \cite{Book} we have argued that the dynamical space has the form of a quantum foam and so non-local quantum effects are to be expected.  So it might be argued that the successful prediction of the masses of these black hole masses, and their dependence on the fine structure constant, is indicative of a grand unification of space and the quantum theory.  This unification is not coming from the quantisation of gravity, but rather from a deeper modelling of reality as an information-theoretic system with emergent quantum-space and quantum matter.

This work is supported by an Australian Research Council Discovery Grant.


\begin{thebibliography}{99}

\bibitem{Book} Cahill  R.T. {\it Process Physics: From Information Theory to Quantum Space
       and Matter},  Nova Science Pub., New York, 2005.
       \bibitem{MMCK} Cahill R.T. and Kitto K. {\it Michelson-Morley Experiments Revisited}, {\it Apeiron}, {\bf 10}(2),104-117, 2003.
        \bibitem{AMGE}  Cahill R.T. {\it Absolute Motion and Gravitational Effects}, {\it Apeiron},   {\bf 11}(1), 53-111, 2004.
  \bibitem{MMC}  Cahill  R.T. {\it The Michelson and Morley 1887 Experiment
and the Discovery of Absolute Motion},   {\it Progress in Physics},  {\bf 3}, 25-29, 2005.
\bibitem{MM}  Michelson A.A. and  Morley E.W. {\it Philos. Mag.} S.5  24 No.151, 449-463, 1887.
\bibitem{Miller}   Miller D.C. {\it Rev. Mod. Phys.},  {\bf 5}, 203-242, 1933.
\bibitem{C5}   Illingworth K.K. {\it  Phys. Rev.} 3,  692-696, 1927.
\bibitem{C6}   Joos G. {\it  Ann. d. Physik} [5] 7,  385, 1930.
\bibitem{C7}   Jaseja T.S. {\it et al.} {\it  Phys. Rev.} A 133, 1221, 1964.
\bibitem{Torr}  Torr D.G. and Kolen P. in  {\it Precision Measurements and Fundamental Constants},  Taylor, B.N. and  Phillips, W.D.  eds.{\it  Natl. Bur. Stand. (U.S.), Spec. Pub.}, 617,  675, 1984.
\bibitem{DeWitte}  Cahill R.T. {\it The Roland DeWitte 1991 Experiment}, {\it Progress in Physics}, {\bf 3}, 60-65, 2006.
\bibitem{Schrod} Cahill R.T. {\it  Dynamical  Fractal  3-Space and the Generalised Schr\"{o}dinger  
Equation: Equivalence Principle and  Vorticity Effects},   {\it Progress in Physics},  {\bf 1}, 27-34, 2006.
\bibitem{alpha}  Cahill R.T.   {\it Gravity, `Dark Matter' and the Fine Structure Constant}, {\it Apeiron}, {\bf
12}(2), 144-177, 2005.
 \bibitem{DM}   Cahill R.T.   {\it  `Dark Matter' as a Quantum Foam In-flow Effect}, in {\it
Trends in Dark Matter Research},  96-140,  ed. J. Val Blain , Nova Science Pub., New York, 2005.    
\bibitem{galaxies}  Cahill   R.T. {\it Black Holes in Elliptical and Spiral Galaxies and in 
Globular Clusters}, {\it Progress in Physics}, {\bf 3}, 51-56, 2005.
\bibitem{boreholes} Cahill  R.T. {\it 3-Space In-flow Theory of Gravity: Boreholes, Blackholes and the Fine Structure Constant},  {\it Progress in Physics}, {\bf 2}, 9-16, 2006.

\bibitem{Crothers} Crothers S.J. {\it A Brief History of Black Holes},  {\it Progress in Physics}, {\bf 2}, 54-57, 2006.

\bibitem{Marconi}  Marconi A. and Hunt L.K. {\it ApJ}, {\bf 589}, L21-L24, part 2, 2003.

\bibitem{Macchetto} Macchetto F. {\it et al.},   {\it ApJ}, {\bf 489}, 579, 1997.

\bibitem{Gebhardt} Gebhardt K. {\it et al.}, {\it  ApJ}, {\bf 583}, 92, 2003.

\bibitem{Bower}  Bower G.A. {\it et al.}, {\it ApJ}, {\bf  550}, 75, 2001.

\bibitem{Maciejewski}  Maciejewski F. and Binney J. {\it MNRAS}, {\bf 323}, 831, 2001.

\bibitem{Verolme} Verolme E.K. {\it et al.}, {\it MNRAS}, {\bf 335}, 517, 2002.

\bibitem{Verdoes} Verdoes Klein G.A., {\it et al.}, {\it AJ}, {\bf 120}, 1221, 2000.

\bibitem{Cappellari}  Cappellari M., {\it et al.}, {\it ApJ}, {\bf 578}, 787, 2002.

\bibitem{Tadhunter} Tadhunter C. {\it et al.},   {\it MNRAS}, {\bf  342},  861, 2003.

\bibitem{Ferrareseb} Ferrarese L. {\it et al.}, {\it ApJ}, {\bf 470},  444, 1996.

\bibitem{Tremaine} Tremaine S. {\it et al.}, {\it ApJ}, {\bf 574}, 740, 2002.

\bibitem {Gebhardtb}  Gebhardt K. {\it et al.}, {\it ApJ}, {\bf 539}, L13,  2000.

\bibitem{Ferrarese} Ferrarese L. and Ford H.C.   {\it ApJ}, {\bf 515},  583, 1999.

\bibitem{Gerssen}  Gerssen J. {\it et al.},  {\it Astron.J.}, {\bf 124}, 3270-3288, 2002; Addendum {\bf 125}, 376, 2003.

\bibitem{Gebhardtc} Gebhardt K. {\it et al.}, {\it ApJ.},  {\bf 578},  L41, 2002.

\bibitem{GPB} Cahill  R.T. {\it Novel Gravity Probe B Frame-Dragging Effect},  {\it Progress in Physics}, {\bf 3}, 30-33, 2005.

\bibitem{Hertz}  Hertz H.  {\it On the Fundamental Equations of Electro-Magnetics for Bodies in Motion}, {\it Wiedemann's Ann.}  {\bf 41}, 369, 1890;  {\it Electric Waves, Collection of Scientific Papers,}  {\it Dover Pub.}, New  York, 1962.

\bibitem{super}  Cahill R.T. {\it The Dynamical Velocity Superposition Effect in the Quantum-Foam In-Flow Theory of Gravity}, physics/0407133.

\bibitem{anisotropy} Cahill R.T. {\it A New Light-Speed Anisotropy Experiment: Absolute Motion and Gravitational Waves Detected},
 {\it Progress in Physics},  {\bf 4}, 73-92, 2006.

\end{thebibliography}
\end{document}